# ■ A *Mathematica* Notebook for Alpha Decay Using an Exponentially Diffuse Boundary Potential

### Eugene F. Chaffin and Daniel S. Banks
**Physics Department, Bob Jones University, 1700 Wade Hampton Blvd., Greenville, SC 29614**

**Abstract** Using the exponentially diffuse boundary potential of Green and Lee (1955), we develop a *Mathematica* notebook to treat alpha decay by using the complex energy approach of Preston (1947), as modified by Pierronne and Marquez (1978). Our approach allows the potential to change slowly from the well depth of the interior of the nucleus to the top of the Coulomb barrier, rather than the sudden step of the simple square well used by Pierronne and Marquez. Recent interest in possible variation in coupling constants such as the strong coupling $\alpha_s$ motivates us to develop this algorithm which can allow numerical study of the variation of the decay constant of a nucleus such as U–238when the depth of the nuclear potential well changes.

**Introduction**
In recent years string theory the dependency of various constants on radii of compactified dimensions has made it appear that half–livesfor alpha decay may have been variable during the early history of the universe [1–3,5,9–11]In order to model this time dependence, we have used *Mathematica* to model the variation of the decay constant with change in depth of the potential well. More than 20 years ago, Pierronne and Marquez [6] treated the theory of alpha decay using a square well solution for the interior of the nucleus and coulombic solutions for the exterior, in a modification of the pioneer work by Preston [8] in which the complex nature of the alpha particle energy is utilized. The Pierronne and Marquez approach enables a finite well depth of around –30 to –100MeV to be used, whereas the earlier approach of Preston required an unrealistic assumption of a positive value for the well depth. The use of the square well leads to spherical Bessel functions for the interior solutions of the square well, which Pierronne and Marquez matched to repulsive coulomb solutions on the boundary of the potential well.

In the 1950's Green and Lee [6] found solutions for a spherical well with an exponentially diffuse boundary potential V(r) = –$V_0$ exp[(a–r)/a$\delta$],which were Bessel functions of nonintegral order for r>a, and the usual spherical Bessel functions for r<a, where a is the radius at which the exponential tail begins and $\delta$ is a dimensionless parameter which characterizes the "shortness" of the tail. We have modified the approach of Pierronne and Marquez to use the Green and Lee solutions. Thus we match the logarithmic derivative of the spherical Bessel function solutions to the nonintegral order Bessel function solutions of Green and Lee at r = a, and then also match these solutions to the Coulombic wave functions at a larger radius r =b.

The Pierronne and Marquez method also requires us to allow the one–bodyalpha particle energy to have a small complex part, thus modifying the bound states to allow tunneling through the barrier. Then we match the imaginary parts of the logarithmic derivative, which leads to the equations giving the decay constant. For the exponentially diffuse boundary wavefunctions of Green and Lee, the index $\nu$ of the Bessel function $J_\nu$ (kx) ) becomes complex. In the limit where the imaginary part of the energy is small, a Taylor series expansion with $\nu$ as the variable may be evaluated at the point where the imaginary part of $\nu$ is zero, leading easily to the expression for the decay constant given in the *Mathematica* code below.

## Results

The algorithm may be used to explore the variation of the decay constant as the well depth, nuclear radius, alpha particle energy at infinity, and other parameters vary. For example, Figure 1 below shows the negative logarithm of the decay constant plotted versus well depth, for a parent nucleus with Z=92, A= 238 (uranium-238).To produce this plot, the inner matching radius a was held constant, while the radius b at which the logarithmic derivatives of the Green and Lee solutions were matched to the coulombic wave functions was determined by Newton-Raphsoniteration, keeping the alpha particle energy fixed at 4.31 MeV.

[Figure 1]

## Conclusions

As has been clearly pointed out by Calmet and Fritsch [1,2], as the strong coupling constant $\alpha_S$ varies, several quantities may vary at once including the nucleon mass. Data such as the natural reactor at Oklo and the Sm-149 cross section do not necessarily constrain these variations if more than one parameter varies at once. Due to the recent observations indicating the cosmological variation of the fine structure constant [10,11], this possibility has to be taken seriously. The algorithm given here may be used to explore the consequences of a variation of $\alpha_S$ on abundances of radioactive nuclides, such as those at Oklo.

## References

[1] Calmet, X. and H. Fritsch, 2001, The cosmological evolution of the nucleon mass and the electroweak coupling constants, hep-ph/0112110.

[2] Calmet, X. and H. Fritsch, 2001, Symmetry breaking and time variation of gauge couplings, hep-ph/0204258.

[3] Damour, T. and F. Dyson, 1996, The Oklo bound on the time variation of the fine-structure constant revisited, *Nucl. Phys*. B 480 , 37-54,hep-ph/9606486.

[4] Fröberg, C.-E.1955. Numerical treatment of coulomb wavefunctions. *Reviews of Modern Physics* 27(4): 399-411.

[5] Fujii, Y. A. Iwamoto, T. Fukahori, T. Ohnuki, M. Nakagawa, H. Hidaka, Y. Oura, and P. Moller, 2000, The nuclear interaction at Oklo 2 billion years ago *Nucl.Phys.* B573, 377-401,hep-ph/9809549.

[6] Green, A. E. S. and K. Lee. 1955. Energy eigenvalues for a spherical well with an exponentially diffuse boundary. *Physical Review* 99(1): 772-777.

[7] Pierronne, M. and L. Marquez. 1978. On the complex energy eigenvalue theory of alpha decay. *Zeitschrift für Physik* A 28619-25.

[8] Preston, M. A. 1947. The theory of alpha-radioactivity.*Physical Review* 71(12): 865-877.

[9] Sisterna, P. and H. Vucetich, 1990, Time variation of fundamental constants: Bounds from gephysical and astronomical data, *Physical Review* D41(2), 1034-1045.

[10] Webb, J.K., V. V. Flambaum, C. W. Churchill, M. J. Drinkwater, and J. D. Barrow,

1999, Search for time variation of the fine structure constant, *Physical Review Letters* 82(5), 884–887, astro-ph/9803165.

[11] Webb, J.K., M.T. Murphy, V.V. Flambaum, V.A. Dzuba, J.D. Barrow, C.W. Churchill, J.X. Prochaska, A.M. Wolfe, 2001, Further evidence for cosmological evolution of the fine structure constant, *Physical Review Letters* 87: 091301, astro-ph/0012539.

## ■ The *Mathematica* Code

```
(*
 An algorithm for finding the decay constant for variable
  well depths. The interior is modeled as a square well with
  an exponentially diffuse edge and a constant potential V_1
  which matches the edge to the exterior Coulomb potential.
   Eugene F. Chaffin and Daniel S. Banks
   May 29, 2002
 The algorithm requires input giving the atomic number Z,
 the atomic mass number A, the kinetic energy of the emitted
  alpha particle E_α and the depth V_0 of the potential well. *)
```

```
(* The distance a is the point where the potential changes from the constant value
    for the interior of the square well to the exponentially diffuse potential V₁-
    V0 Exp[(a-x)/(a δ)], in units of 10*10⁻¹⁵ meters.
    The distance b is the point where the potential changes from the exponentially
    diffuse potential to the Coulomb potential, in units of 10*10⁻¹⁵ meters. *)
a = 1;
b = 1.045;
V0 = 50;
V₁ = (8.98755*10⁹ * 2 * 90 * (1.6021*10⁻¹⁹)² / (b*10⁻¹⁴)) / (1.6021*10⁻¹³) +
    V0 Exp[(a - b) / (a δ)];
δ = 0.035;
M = 6.645 10⁻²⁷
Z = 90
AA = 234
E_α = 4.31
K_ex = ((2 * M * E_α * 1.6021 10⁻¹³).⁵) / (1.0546 10⁻³⁴);
Print["K_ex = ", K_ex]
R = b*10⁻¹⁴
E₀ = (1.05457 * 10⁻³⁴)² / (2 * 6.4 * 10⁻²⁷ * a² * 10⁻²⁸) / (1.6021 * 10⁻¹³)
ϵ₀ = √(V0/E₀) ;
k = 2 * δ * ϵ₀;
RR = R / (10⁻¹⁴);
v = e^((1-RR/a)/(2*δ));
```

$6.645 \times 10^{-27}$

$90$

$234$

$4.31$

$K_{ex} = 9.0836 \times 10^{14}$

$1.045 \times 10^{-14}$

$0.0542315$

This portion of the notebook provides an iterative process for finding the order n (non–integral) for the Bessel function which is the solution connecting the spherical Bessel function to the Coulomb function. The user provides an initial guess for n and calculates `-2 * δ * (ϵ' / Tan[ϵ']) + 2 * δ * ϵ₀ * BesselJ[n + 1, k] / BesselJ[n, k]`. When the result is the same or close enough to the original value, the process is finished.

```mathematica
n = 1.60726
ϵ_w = n / (2 * δ)
ϵ' = Sqrt[ϵ_0^2 - ϵ_w^2] + V_1
k = 2 * δ * ϵ_0
```

1.60726

22.9609

20.8182

2.12548

```mathematica
-2 * δ * (ϵ' / Tan[ϵ']) + 2 * δ * ϵ_0 * BesselJ[n + 1, k] / BesselJ[n, k]
```

1.60746

```mathematica
η = 0.6300 * Z * ((AA / (E_α * (AA + 4)))^.5);
GR[T_] :=
 Exp[-2 * η * ((T * (1 - T))^.5 + ArcSin[T^0.5] - π / 2) + 0.25 * Log[T / (1 - T)] +
    (8 * T^2 - 12 * T + 9) / (48 * Sqrt[T] * (1 - T)^1.5) / (2 * η) +
    (8 * T - 3) / (64 * T * (1 - T)^3) / ((2 * η)^2) -
    (2048 * T^6 - 9216 * T^5 + 16128 * T^4 - 13440 * T^3 -
        12240 * T^2 + 7560 * T - 1890) / (92160 * T^1.5 * (1 - T)^4.5) / ((2 * η)^3) +
    3 * (1024 * T^3 - 448 * T^2 + 208 * T - 39) / (8192 * (T^2) * (1 - T)^6) /
       ((2 * η)^4) -
    (-(262144 * T^10 - 1966080 * T^9 + 6389760 * T^8 - 11714560 * T^7) -
        (13178880 * T^6 - 9225216 * T^5 + 13520640 * T^4) -
        (-3588480 * T^3 + 2487240 * T^2 - 873180 * T + 130977)) / (10321920 * (T^2.5) * (1 - T)^7.5) /
     ((2 * η)^5) + ((1105920 * T^5 - 55296 * T^4 + 314624 * T^3 - 159552 * T^2) +
        (45576 * T - 5697)) / (393216 * (T^3) * (1 - T)^9) / ((2 * η)^6)];
GRD[T_] := GR'[T];
```

```mathematica
FR[T_] :=
  .5 * Exp[ 2 * η * ((T * (1 - T))^.5 + ArcSin[T^.5] - π / 2) +
     0.25 * Log[T / (1 - T)] -
      ((8 * T^2 - 12 * T + 9) / (48 * Sqrt[T] * (1 - T)^1.5)) /
        (2 * η) + ((8 * T - 3) / (64 * T * (1 - T)^3)) / ((2 * η)^2) +
      ((2048 * T^6 - 9216 * T^5 + 16128 * T^4 - 13440 * T^3 - 12240 * T^2 +
             7560 * T - 1890) / (92160 * T^1.5 * (1 - T)^4.5)) /
        ((2 * η)^3) + (3 * (1024 * T^3 - 448 * T^2 + 208 * T - 39) /
           (8192 * (T^2) * (1 - T)^6)) / ((2 * η)^4) +
      ((-(262144 * T^10 - 1966080 * T^9 + 6389760 * T^8 -
                11714560 * T^7) -
            (13178880 * T^6 - 9225216 * T^5 + 13520640 * T^4) -
            (-3588480 * T^3 + 2487240 * T^2 - 873180 * T + 130977)) /
         (10321920 * (T^2.5) * (1 - T)^7.5)) / ((2 * η)^5) +
      (((1105920 * T^5 - 55296 * T^4 + 314624 * T^3 - 159552 * T^2) +
            (45576 * T - 5697)) /
         (393216 * (T^3) * (1 - T)^9)) / ((2 * η)^6)];

FRD[T_] := FR'[T];

(* Having found n in the above procedure,
 we now start a loop for Newton-Raphson iteration to find the
   radius R by matching the logarithmic derivatives for the non-
  integral order Bessel function and the Coulomb wavefunction *)
TES =
  1;
```

```mathematica
While[TES > 1/10^20,
 ρ_e = K_ex * R;
 RR = R / (10^-14);
 v = e^((1-RR/a)/(2*δ));
 T = ρ_e / (2 * η);
 (* When T is greater than 1,
    the Riccati method of expanding the Coulomb wavefunction is invalid
    (Froberg, 1955), so an error message is printed in that case *)
 If[T > 1, Print["T = ", T]];
 FOP = FRD[T] / (2 * η);
 GOP = GRD[T] / (2 * η);
 (* Find the Wronskian as a check on the accuracy *)
 W = GR[T] * FOP - FR[T] * GOP;
 If[Abs[1 - W] > 10^-4, Print["W = ", W]];
 GOPP = -GR[T] * (1 - η * 2 / ρ_e);
 FN = K_ex * GOP / GR[T] +
    1 / (2 * δ * a * 10^-14) * (n - k * v * BesselJ[n + 1, k * v] / BesselJ[n, k * v]);
 FNP = (K_ex * GOPP / GR[T] - K_ex * GOP * GOP / (GR[T]^2)) * K_ex +
    (1 / (2 * δ * a * 10^-14)) *
     ((k^2 * v^2 / (2 * δ * a * 10^-14 * BesselJ[n, k * v])) * ((2 / (k * v)) * BesselJ[n + 1, k * v] +
        (BesselJ[n + 1, k * v])^2 / BesselJ[n, k * v] - BesselJ[n + 2, k * v]));
 RP = R - FN / FNP;
 TES = Abs[RP - R];
 R = RP;]
(* end of Newton-Raphson iteration loop *)

Print["R = ", R]
```

R = $1.04499 \times 10^{-14}$

```mathematica
B = (K_ex * RP * (GR[T] * GOPP - GOP^2) + GR[T] * GOP) / (GR[T]^2);

CC = B * ((2 * M / (E_α * 1.6021 10^-13))^.5) / (1.0546 10^-34) / 4;

x = k * e^((1-R/(a*10^-14))/(2*δ));

ν = n;
```

$$Q = e^{(1-R/(a*10^{-14}))/2*\delta} * \left(1/x + \left(\frac{\frac{\nu}{x} * \text{BesselJ}[\nu, x] - \text{BesselJ}[\nu+1, x]}{(\text{BesselJ}[\nu, x])^2} - \frac{1}{x * \text{BesselJ}[\nu, x]}\right) * \right.$$

$$\left(\frac{2^{-\nu} * \text{PolyGamma}[\nu+1] * x^{\nu}}{\text{Gamma}[\nu+1]} - \frac{2^{-\nu-2} * \text{PolyGamma}[\nu+2] * x^{\nu+2}}{\text{Gamma}[\nu+2]} + \right.$$

$$\frac{2^{-\nu-4} * \text{PolyGamma}[\nu+3] * x^{\nu+4}}{\text{Gamma}[\nu+3]} - \frac{2^{-\nu-6} * \text{PolyGamma}[\nu+4] * x^{\nu+6}}{\text{Gamma}[\nu+4]} +$$

$$\frac{2^{-\nu-8} * \text{PolyGamma}[\nu+5] * x^{\nu+8}}{\text{Gamma}[\nu+5]} - \frac{2^{-\nu-10} * \text{PolyGamma}[\nu+6] * x^{\nu+10}}{\text{Gamma}[\nu+6]} +$$

$$\frac{2^{-\nu-12} * \text{PolyGamma}[\nu+7] * x^{\nu+12}}{\text{Gamma}[\nu+7]} - \frac{2^{-\nu-14} * \text{PolyGamma}[\nu+8] * x^{\nu+14}}{\text{Gamma}[\nu+8]} +$$

$$\frac{2^{-\nu-16} * \text{PolyGamma}[\nu+9] * x^{\nu+16}}{\text{Gamma}[\nu+9]} - \frac{2^{-\nu-18} * \text{PolyGamma}[\nu+10] * x^{\nu+18}}{\text{Gamma}[\nu+10]} +$$

$$\frac{2^{-\nu-20} * \text{PolyGamma}[\nu+11] * x^{\nu+20}}{\text{Gamma}[\nu+11]} - \frac{2^{-\nu-22} * \text{PolyGamma}[\nu+12] * x^{\nu+22}}{\text{Gamma}[\nu+12]} +$$

$$\frac{2^{-\nu-24} * \text{PolyGamma}[\nu+13] * x^{\nu+24}}{\text{Gamma}[\nu+13]} - \frac{2^{-\nu-26} * \text{PolyGamma}[\nu+14] * x^{\nu+26}}{\text{Gamma}[\nu+14]} +$$

$$\left.\left.\frac{2^{-\nu-28} * \text{PolyGamma}[\nu+15] * x^{\nu+28}}{\text{Gamma}[\nu+15]}\right)\right);$$

**(* Calculate the decay constant *)**

$$\lambda = \left(K_{ex} \Big/ \left(\left(CC + \frac{k*\delta*M*a*10^{-14}}{(1.0546\ 10^{-34})^2} * Q\right) * (1.0546\ 10^{-34})\right)\right) \Big/$$
$$(GR[T] * GR[T]);$$

`Print["λ = ", λ]`

$\lambda = 1.6956 \times 10^{-13}$

`V₁`

38.6247

$\left(8.98755 * 10^9 * 2 * 90 * (1.6021 * 10^{-19})^2 \big/ (b * 10^{-14})\right) \big/ (1.6021 * 10^{-13})$

24.802

**(* Print the radius resulting from the matching *)**
`Print["R = ", R]`

$R = 1.04499 \times 10^{-14}$

`Q`

1.12317

`CC`

$-2.78579 \times 10^{26}$

```
k
```

2.12548

```
M
```

$6.645 \times 10^{-27}$

```
δ
```

0.035

```
a
```

$$\epsilon' = \sqrt{{\epsilon_0}^2 - {\epsilon_w}^2 + V_1}$$

20.8182

```
h[ρ_] := ((ρ)^0.5) BesselJ[0.5, ε' * ρ];
```

```
h[1]
```

0.161212

```
BesselJ[n, k]
```

0.482814

```
h[1] / BesselJ[n, k]
```

0.333902

$$\int_0^1 \left( ((\rho)^{0.5}) \, \text{BesselJ}[0.5, \epsilon' * \rho] \right)^2 \, d\rho$$

0.0155523

$$N\left[ \int_1^{10} (0.334227)^2 \left( \text{BesselJ}[n, k * E^{(1-\rho)/(2*\delta)}] \right)^2 \, d\rho \right]$$

0.000842111

$$1 / \sqrt{0.015553 + .000842111}$$

7.80985

```
7.80985 * (0.334227)
```

2.61026

```
GR[T_] := (8.51346*10^-12) (2/(a*10^-14))^0.5 * ((2*η*T)/K_ex) *
  Exp[-2*η*(((T*(1-T))^.5 + ArcSin[T^0.5] - π/2) + 0.25*Log[T/(1-T)] +
    (8*T^2 - 12*T + 9) / (48*Sqrt[T] * (1-T)^1.5) / (2*η) +
    (8*T - 3) / (64*T*(1-T)^3) / ((2*η)^2) -
    (2048*T^6 - 9216*T^5 + 16128*T^4 - 13440*T^3 -
      12240*T^2 + 7560*T - 1890) / (92160*T^1.5 * (1-T)^4.5) / ((2*η)^3) +
    3*(1024*T^3 - 448*T^2 + 208*T - 39) / (8192*(T^2)*(1-T)^6) /
      ((2*η)^4) -
    (-(262144*T^10 - 1966080*T^9 + 6389760*T^8 - 11714560*T^7) -
      (13178880*T^6 - 9225216*T^5 + 13520640*T^4) -
      (-3588480*T^3 + 2487240*T^2 - 873180*T + 130977)) / (10321920*(T^2.5)*(1-T)^7.5) /
    ((2*η)^5) + ((1105920*T^5 - 55296*T^4 + 314624*T^3 - 159552*T^2) +
      (45576*T - 5697)) / (393216*(T^3)*(1-T)^9) / ((2*η)^6)]

TR = R*K_ex / (2*η)
```

0.175257

```
GR[TR]
```

0.603691

```
gt[ρ_] := 2.61006*BesselJ[n, k*E^((1-ρ)/(2*δ))] / ρ;

ρR = R / 10^-14
```

General::spell1 : Possible spelling error: new symbol name "ρR" is similar to existing symbol "ρ".

1.04499

```
gt[ρR]
```

0.603773

```
0.603888 / (7.09333*10^10)
```

$8.51346 \times 10^{-12}$

```mathematica
g[ρ_] := Which[ρ < 1, (7.80923) * ((ρ)^0.5) BesselJ[0.5, ϵ' *ρ], 1 ≤ ρ ≤ (R / 10^-14),
    2.61006 * BesselJ[n, k * E^((1-ρ)/(2*δ))] / ρ, (R / (a * 10^-14)) ≤ ρ, GR[K_ex ρ * a * 10^-14 / (2 * η)]];
fV[ρ_] := Which[ρ < 1, -V0,
   1 ≤ ρ ≤ b, V_1 - V0 Exp[(1 - ρ) / (δ)], ρ > b,
   (8.98755 * 10^9 * 2 * 90 * (1.6021 * 10^-19)^2 / (ρ * 10^-14)) / (1.6021 * 10^-13)];
Plot[{fV[ρ], 10 * g[ρ]}, {ρ, 0.01, 2}, PlotRange → All, PlotLabel → "\!\(\*
StyleBox[\"Wavefunction\",\nFontFamily->\"Arial\",\nFontSize->24]\)\!\(\*
StyleBox[\"×\",\nFontFamily->\"Arial\",\nFontSize->24]\)\!\(\*
StyleBox[\"10\",\nFontFamily->\"Arial\",\nFontSize->24]\)\!\(\*
StyleBox[\" \",\nFontFamily->\"Arial\",\nFontSize->24]\)\!\(\*
StyleBox[\"and\",\nFontFamily->\"Arial\",\nFontSize->24]\)\!\(\*
StyleBox[\" \",\nFontFamily->\"Arial\",\nFontSize->24]\)\!\(\*
StyleBox[\"Potential\",\nFontFamily->\"Arial\",\nFontSize->24]\)\!\(\*
StyleBox[\" \",\nFontFamily->\"Arial\",\nFontSize->24]\)\!\(\*
StyleBox[\"versus\",\nFontFamily->\"Arial\",\nFontSize->24]\)\!\(\*
StyleBox[\" \",\nFontFamily->\"Arial\",\nFontSize->24]\)\!\(\*
StyleBox[\"Distance\",\nFontFamily->\"Arial\",\nFontSize->24]\)",
 AxesLabel → {ρ , psi[V]}, PlotStyle → {RGBColor[1, 0, 0], RGBColor[0, 1, 0]},
 TextStyle → {FontSize → 24}];

W

1.
```